\documentclass[12pt,preprint]{aastex}
\def\kms{{\rm km\,s^{-1}}}
\def\kpc{{\rm kpc}}

\def\lim{{\rm lim}}

\def\bv{{\bf v}}
\newcommand{\bdv}[1]{\mbox{\boldmath$#1$}}
\def\e{{\rm E}}

\def\rel{{\rm rel}}
\def\pirel{\pi_{\rm rel}}
\def\au{{\rm AU}}

\def\bmu{{\bdv{\mu}}}
\def\btheta{{\bdv{\theta}}}
\def\kms{{\rm km}\,{\rm s}^{-1}}

\begin{document}

\title{Probing MACHOs Toward the Galactic Bulge}

\author{Andrew Gould}
\affil{Department of Astronomy, The Ohio State University,
140 W.\ 18th Ave., Columbus, OH 43210}
\authoremail
{gould@astronomy.ohio-state.edu}

\singlespace

\begin{abstract}

If the massive compact halo object (MACHO) fraction of the 
Galactic dark halo is $f\sim 20\%$ as suggested by some microlensing
experiments, then about $1.2\%$ of lensing events toward the
Galactic bulge are due to MACHOs.  For the 40\% of these that lie
nearby ($D_l<4\,\kpc$), measurement of their distance $D_l$ would
distinguish them from bulge lenses, while measurement of their
transverse velocity $\bv_l$ would distinguish them from disk lenses.
Hence, it would be possible to identify about $0.5\%(f/20\%)$ of
all events as due to MACHOs.  I show that a planned experiment
using the {\it Space Interferometry Mission (SIM PlanetQuest)}
could thereby detect 1 or 2 such events.  This is at the margin
of what is required because of a small, but non-negligible background
from spheroid stars.

\end{abstract}
 
\section{Introduction
\label{sec:intro}}

Following the suggestion of \citet{pac86}, the MACHO \citep{alcock93}
and EROS \citep{aubourg93} collaborations began searching for dark matter
in the form of massive compact halo objects (MACHOs) by microlensing
observations toward the Large Magellanic Cloud (LMC).  This target
seemed ideal because of the small column of known populations of
stars compared to the
huge volume of space that would be home to the putative MACHOs.
The microlensing optical depth due to known stars was estimated
to be $\tau_{\rm MW}^{\rm LMC}=8\times 10^{-9}$ for the Milky Way 
disk \citep{gbf97}
and $\tau_{\rm LMC}^{\rm LMC}= 1\times 10^{-8}$ for the LMC itself 
\citep{gould95b}.
By contrast, if the dark halo were completely composed of MACHOs,
their optical depth would be of order,
\begin{equation}
\tau^{\rm LMC}_{\rm halo}\sim {v_{\rm rot}^2\over c^2} = 5\times 10^{-7},
\label{eqn:taulmchalo}
\end{equation}
roughly 25 times higher.  Here, $v_{\rm rot}=220\,\kms$ is the Milky
Way rotation speed.  Hence, when the experiments began, it seemed
as though even a crude measurement of $\tau$ would unambiguously
determine whether the halo was composed of MACHOs.

A decade later, the situation is far less clear than was anticipated.  MACHO 
\citep{alcock00} found $\tau\sim 1\times 10^{-7}$, roughly the root-mean-square
of the results expected from MACHOs and stars.  They interpreted this
to mean that the halo was 20\% composed of MACHOs and estimated the typical
mass to be $M\sim 0.4\,M_\odot$.  On the other hand, the EROS collaboration
\citep{afonso03a,tisserand05} found an upper limit for the optical depth
due to MACHOs of 5\% of the full-halo value.

One option for resolving this conflict is to explore other lines of sight.
\citet{crotts92} and \citet{baillon93} advocated M31, and several
collaborations, including AGAPE \citep{ansari99}, Columbia-VATT 
\citep{uglesich04}, MEGA \citep{dejong04}, NainiTal \citep{joshi05}
POINT-AGAPE \citep{auriere01}, SLOTT-AGAPE \citep{calchi03}, and
WeCAPP \citep{riffeser03}, have pursued this suggestion.  In many
ways this is substantially more challenging than the observations
toward the LMC, simply because M31 is 15 times farther away and
hence the sources are substantially fainter.  Events are now being
reported from these experiments, and their implications for dark matter
should be available soon.

The microlensing target field that has been monitored the most intensively
is the Galactic bulge.  Originally proposed by \citet{pac91} and
\citet{griest91}, major surveys have been carried out by the
OGLE \citep{udalski93,udalski03}, DUO \citep{alard95}, MACHO \citep{popow05},
EROS \citep{afonso03b}, and MOA \citep{abe04} collaborations.
The primary motivation of both proposals was to probe for disk dark matter
and other exotic objects such as a large population of Jupiters.
\citet{griest91} does mention that if the halo is composed of MACHOs, then
these will give rise to an optical depth 
$\tau_{\rm halo}^{\rm bulge}=1.3\times 10^{-7}$, but since this is 4 times
smaller than the predicted optical depth due to disk stars
$\tau_{\rm disk}^{\rm bulge}=5.1\times 10^{-7}$, there did not appear
to be any way to isolate the MACHO events.

Bulge microlensing observations have been enormously fruitful.
\citet{kiraga94} showed that the optical depth due to bulge self-lensing
was even greater than that due to disk stars.  The high event rate
encouraged searches for lensing anomalies due to planetary 
companions of the lenses \citep{mao91,gould92,rhie00,albrow01b,gaudi02,abe04},
which has now yielded the first firm microlensing planet detection
\citep{bond04}.  Bulge microlensing has enabled the 
first microlens mass measurement \citep{an02} and the
probing of bulge-star atmospheres with $\mu$as resolution both photometrically
\citep{alcock97,albrow99,albrow00,fields03} and spectroscopically
\citep{castro01,albrow01a,cassan04}.

Here I show that bulge microlensing can also be used to probe for 
halo dark matter (MACHOs) in the inner Galaxy.  This seems absurd at
first sight because the observed optical depth, 
$\tau_{\rm bulge}^{\rm obs}\sim 2\times 10^{-6}$, is about 15 times
higher than the rate predicted by \citet{griest91},
$\tau_{\rm bulge}^{\rm halo}\sim 1.3\times 10^{-7}$, even assuming that
the dark halo were completely composed of MACHOs.  However, the microlensing
experiments toward the LMC seem to imply that this fraction is no larger
than 20\%, which means that only about 1\% of Galactic bulge microlensing
would be due to halo objects.  How would one identify these halo
microlensing events within the barrage of microlensing by ordinary
bulge and disk stars?

\section{Needle in Haystack
\label{sec:haystack}}

Halo lenses are distinguished from disk lenses by the their transverse
velocity $\bv_l$ 
relative to the Sun, and from bulge lenses by their distance
from the Sun, $D_l$ (or equivalently, their absolute parallax $\pi_l$).
Hence, to reliably identify the nearby, fast MACHOs, one must
reliably measure $\bv_l$ and $\pi_l$.  Since the MACHOs are by definition
``dark'' matter, direct observations of the lens cannot be employed
in making these determinations, as they were for example for MACHO-LMC-5
\citep{alcock01,dck04,gould04a,gba04}.  Instead, these quantities must
be derived entirely from observations of the source during and after the
microlensing events.

\subsection{Observables
\label{sec:observables}}

These two quantities can be expressed in terms of microlensing observables
by (e.g., \citealt{gould00}),
\begin{equation}
\pi_l = \pirel + \pi_s,\qquad \pirel =\pi_\e\theta_\e
\label{eqn:pil}
\end{equation}
and
\begin{equation}
\bv_l = {\bmu_\rel + \bmu_s\over \pirel + \pi_s}\au,\qquad
\bmu_\rel = {\btheta_\e\over t_\e}.
\label{eqn:bvl}
\end{equation}
Here, $\pi_l$, $\pi_s$ $\bmu_l$, $\bmu_s$ are the absolute parallaxes and
proper motions of the lens and source, $\pi_\rel=\pi_l-\pi_s$ and
$\bmu_\rel=\bmu_l-\bmu_s$ are the lens-source relative parallax and
proper motion, $\theta_\e$ is the angular Einstein radius, $t_\e$ is
the Einstein timescale, and $\pi_\e$ is the microlens parallax
(i.e., the inverse of the projected Einstein radius, 
$\pi_\e = \au/\tilde r_\e$).  The direction of $\btheta_\e$ is that of
the lens-source relative proper motion.

In brief, to determine $\pi_l$ and $\bv_l$, one must measure five observables,
two 2-vectors ($\bmu_s$ and $\btheta_\e$) and three scalars 
($\pi_s, \pi_\e,$ and $t_\e$).

\subsection{Parameter Measurement
\label{sec:parms}}

Two of these five parameters ($\pi_s$ and $\bmu_s$) are
related solely to the source, while the remaining three ($\pi_\e$, $t_\e$,
and $\btheta_\e$) are microlensing-event parameters.  Of these three, only
one $(t_\e)$ is routinely measured during microlensing events.  The other
two are higher order parameters.  While there are a variety of 
methods to measure $\pi_\e$ and $\theta_\e$ (see \citealt{gould01}),
these generally apply to only a small fraction of events.  There
are only two events (out of almost 3000 discovered) for which
both parameters have been measured from microlensing data alone 
\citep{an02,kubas05}, and both of these were binary lenses.

The only known way to {\it routinely} determine $\btheta_\e$
is by high-precision astrometric measurements of the microlensing
event \citep{HNP95,MY95,Wa95,Pa98,BSV98}.  The centroid of the microlensed
images deviates from the source position by an amount and direction
that yields both components of $\btheta_\e$.

The only known way to {\it routinely} determine $\pi_\e$ is to
make photometric measurements of the event from two locations separated
by of order $\tilde r_\e$ \citep{Re66,gould94}.  The difference in the
event parameters then yields both the size of $\tilde r_\e$ and the 
direction of motion (the latter potentially confirming the direction
extracted from $\btheta_\e$).
Since $\tilde r_\e \sim O(\au)$, in practice this means placing a satellite
in solar orbit.  Although there is a four-fold ambiguity in the determination
of $\pi_\e$,
this can be resolved by higher-order effects \citep{gould95a}.  Moreover,
measurement of the direction of $\btheta_\e$ also helps resolve this
degeneracy.

\subsection{{\it SIM PlanetQuest} Measurements
\label{sec:sim}}

\citet{gs99} showed that the 
{\it Space Interferometry Mission (SIM PlanetQuest)} combined with
ground-based photometry,  could determine both
of these parameters with good $\sim 3\%$ precision with about 5 hours total 
obsrvation time for bright $(I\sim 15)$
events having typical lens parameters.  
Moreover, they showed that the same observations would
also yield good measurements of $\pi_s$ and $\bmu_s$.  Hence, {\it SIM}
(combined with ground-based photometry) could measure all the required
quantities for about 200 events with about 1000 hours of observing time.
Indeed, a {\it SIM} Key Project has been awarded 1200 hours of observation
time to carry out such observations.  The main objective of this project
is to measure the bulge mass function but the same observations could
cull out the handful of halo events that could be present in the same
sample.

{\it SIM} has been descoped since \citet{gs99} made their analysis.
The new performance is not precisely known but it is likely that the
precision will degrade to something like $\sim 5\%$ for $\pi_\e$ and
$\sim 10\%$ for $\theta_\e$ for the canonical events considered by
\citet{gs99}.  Moreover, it is unlikely that 200 $I=15$ events will
be found during the 5-year primary {\it SIM} mission, and using
fainter sources (e.g., $I=16.5$) would further degrade the
precision by a factor 2.  Nevertheless, as I show below, this
precision would be quite adequate for distinguishing halo lenses.

\section{Background from Spheroid/Bulge Stars
\label{sec:spheroid}}

Halo lenses could produce events anywhere along the line of sight
from the Sun to the bulge and, assuming an isothermal halo model
with core radius $a=5\,\kpc$, the density
\begin{equation}
\rho_{\rm halo} = {v_c^2\over 4\pi G(R^2 + a^2)}
\label{eqn:rhoh}
\end{equation}
rises all the way to the Galactic center.  Here $R$ is Galactocentric
distance.  The optical depth per unit path length along a line of
sight toward the Galactic center therefore also rises almost all
the way in,
\begin{equation}
{d\tau_{\rm halo}\over d D_l} = {\rho_{\rm halo}\over M}\,
{\pi r_\e^2}  = {v_c^2\over c^2}\,{f\over R_0}
\,{x(1-x)\over (a/R_0)^2+(1-x)^2}.
\label{eqn:tauhalo}
\end{equation}
Here, $r_\e=(4GM D_{l} D_{s}/c^2 D_{os})^{1/2}$ is the Einstein
radius, $D_l$ and $D_s$ are the source and lens distances, $D_{ls}=D_s-D_l$,
$f$ is the fraction of the halo in the form of MACHOs, $R_0=8\,\kpc$ is the
Solar Galactocentric distance, $x\equiv D_l/R_0$, 
and $M$ (which cancels out) is the mass of the lens.

However, in the inner Galaxy, these halo lenses are completely 
submerged in the background of bulge lenses, and since they
have similar kinematics, there is no way to reliably distinguish them.
It is only out closer to the Sun, where the spheroidal population
(here usually called ``spheroid'' or ``stellar halo''), thins out that
one may hope to separate the two populations.
Even here, there is some possibility of contamination.  The
local spheroid density is only about 1\% of the dark halo, but
if $f\sim 20\%$ as \citet{alcock00} suggest, then MACHOs are only
20 times more common than spheroid stars locally.  Moreover, 
as one approaches the Galactic center, the spheroid density
grows substantially more rapidly than does the dark halo.  To make
a quantitative comparison, I adopt 
\begin{equation}
\rho_{\rm spheroid} = 1\times 10^{-4}{M_\odot\over \rm pc^3}
\biggl({R\over R_0}\biggr)^{-3.2}.
\label{eqn:rhospheroid}
\end{equation}
After accounting for observed stars and extrapolating down to
brown dwarfs and up to the progenitors of remnants, \citet{gfb98} 
estimate $6.4\times 10^{-5}M_\odot\,\rm pc^{-3}$.  
However, both \citet{dahn95} and
\citet{gould03} find substantially more low-luminosity ($M_V>8$) stars
than did \citet{gfb98} in their more local sample
(see Fig.~2 from \citealt{gould04b}), so I have adjusted
their estimate upward.
The power-law slope is measured by several techniques (\citealt{gfb98} and
references therein).  

\begin{figure}
\plotone{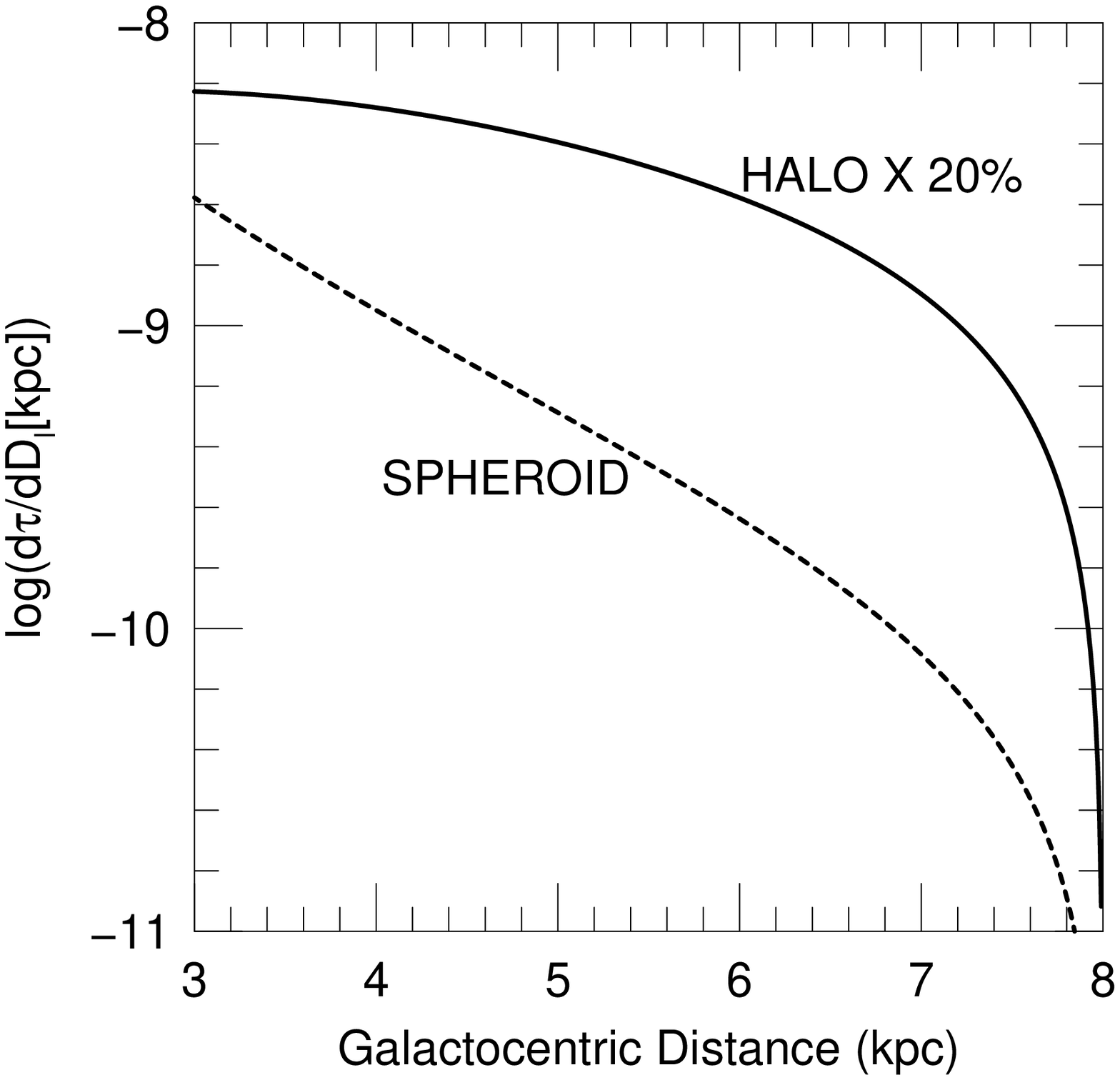}
\caption{\label{fig:opdep}
Optical Depth per unit path length $d\tau/dD_l$ as a function of
distance from the Galactic center for a source near the Galactic
center.  The halo (assuming a $f=20\%$ MACHO fraction) and the spheroid
are shown by {\it solid} and {\it dashed} curves, respectively.  
For $f=20\%$, spheroid stars are a 20\% background at $R=4\,\kpc$ and
a 10\% background at $R=7\,\kpc$, which implies that 2 or 3 halo lenses
must be identified at $R>4\,\kpc$ for a reliable halo ``detection''.
Inside $R<4\,\kpc$ the spheroid continues to grow (and also transforms
into the bulge), making the identification of halo lenses less secure.
Hence, the experiment should be restricted to $R>4\,\kpc$, where the
total optical depth is $\tau = 5\times 10^{-8}f$.
}\end{figure}

Figure \ref{fig:opdep} shows the optical depth per unit distance due 
to spheroid stars and to 
putative MACHOs under the assumption that $f=20\%$. 
It shows that even with a MACHO fraction of 20\%, the halo dominates
the spheroid from $R=R_0$ to  
$R=4\,\kpc$, which latter is about the limit to which the
local spheroid density profile can be reliably extrapolated.  However,
this domination is not overwhelming: at $R=4\,\kpc$ it is only a
factor of 5 and even at $R=7\,\kpc$ (where the halo optical depth
has fallen by a factor 5) the halo only dominates by a factor 10.
This means that 2 or 3 halo lenses would have to be identified to
constitute a reliable ``MACHO detection''.  Otherwise, there would
be a significant possibility that spheroid lenses were responsible.

Since one must restrict attention to $D_l<4\,\kpc$, the total available
halo optical depth is reduced by a factor $0.4$ relative to the
$1.3\times 10^{-7}$ calculated by \citet{griest91}.  If we further assume
$f=20\%$, the available halo optical depth is further reduced to $10^{-8}$,
about 0.5\% of the observed optical depth of 
$\tau\sim 2\times 10^{-6}$ \citep{afonso03b,popow05,sumi05}.  
Hence, assuming for the
moment that the event rates are in proportion to the optical depths,
roughly 200 measurements would be required to identify a single halo lens.
Thus, if the {\it SIM} mission were extended from 5 to 10 years (as is
currently envisioned) then one might expect to find about 2 halo lenses.
As noted above this is just at the margin of a viable detection.

\section{Practical Considerations
\label{sec:practical}}

\subsection{Event Timescales
\label{sec:timescale}}

The Einstein timescales of these halo events are given by,
\begin{equation}
t_\e = 15\,{\rm day}
\biggl[{M\over 0.4\,M_\odot}\,{x(1-x)\over 0.25}\biggr]^{1/2}
\biggl({v_\perp\over 300\,\kms}\biggr)^{-1},
\label{eqn:teeval}
\end{equation}
where $v_\perp$ is the transverse lens velocity relative to the
observer-source line of sight.
Thus, for the mass range advocated by \citet{alcock00}, the
typical event timescales will be fairly short.  This is important
because the event must be identified and alerted to the satellite
well before peak in order to measure $\pi_\e$ \citep{gs99}.
Hence, a fairly aggressive posture is required to keep the halo
events in the sample.

However, the fact that these halo events are somewhat shorter than 
typical bulge events means that they are also more frequent than
would be indicated by their optical depth alone.  That is, the
event rate $\Gamma\propto \tau/t_\e$, so the rate is inversely
proportional to the timescale.  Hence, the shorter timescales
enhances the viability of a given experiment relative to what
was discussed in \S~\ref{sec:spheroid}, provided that not too
many halo events are lost because they are too short.

\subsection{Signal-to-Noise Ratios
\label{sec:sn}}

The two microlensing parameters being measured are related to the
underlying physical parameters by,
\begin{equation}
\pi_\e = \sqrt{\pi_\rel \over \kappa M},\qquad
\theta_\e = \sqrt{\kappa M\pi_\rel},
\label{eqn:piethetae}
\end{equation}
where $\kappa = 4G/\au c^2 \sim 8.1\,{\rm mas}\,M^{-1}_\odot$.
Hence, for fixed $M$, both $\pi_\e$ and $\theta_\e$ 
are proportional to $\pi_\rel^{1/2}$.
Since the absolute errors in these two quantities are approximately
independent of their size, this means that the fractional errors
decline as $\pi_\rel^{-1/2}$.  The basic experiment is designed
for typical bulge-bulge lensing, in which the lenses are of order
$M\sim 0.5\,M_\odot$ and the relative parallaxes are 
$\pi_\rel\sim \au/7\,\kpc - \au/9\,\kpc =31\,\mu$as. By contrast,
since $D_l\leq 4\,\kpc$, the halo-lens relative parallaxes are
$\pi_\rel>125\,\mu$as.
Hence, if a halo event is successfully monitored, both $\pi_\e$ and
$\theta_\e$ will be measured substantially more accurately than for
typical events.

\acknowledgments
This work was supported by grant AST 02-01266 from the NSF and by
JPL contract 1226901.

\clearpage

\clearpage

\begin{thebibliography}{}

\bibitem[Abe et al.(2004)]{abe04} Abe, F. et al. 2004, Science, 305, 1264

\bibitem[Afonso et al.(2003a)]{afonso03a} Afonso, C. et al. 2003a, \aap, 400, 
951

\bibitem[Afonso et al.(2003b)]{afonso03b} Afonso, C. et al. 2003b, \aap, 404, 
145 

\bibitem[Alard et al.(1995)]{alard95} Alard, C., Mao, S., \& Guibert, J. 1995
\aap, 300, L17

\bibitem[Albrow et al.(1999)]{albrow99} Albrow, M.D. et al. 1999, 
\apj, 512, 672

\bibitem[Albrow et al.(2000)]{albrow00} Albrow, M.D. et al. 2000, 
\apj, 534, 894

\bibitem[Albrow et al.(2001a)]{albrow01a} Albrow, M.D. et al. 2001a, 
\apj, 550, L173

\bibitem[Albrow et al.(2001b)]{albrow01b} Albrow, M.D. et al. 2001b, 
\apj, 556, L113

\bibitem[Alcock et al.(1993)]{alcock93} Alcock, C. et al. 1993, Nature, 365, 621

\bibitem[Alcock et al.(1997)]{alcock97} Alcock, C. et al. 1997, \apj, 491, 436

\bibitem[Alcock et al.(2000)]{alcock00} Alcock, C. et al. 2000, \apj, 542, 281

\bibitem[Alcock et al.(2001)]{alcock01} Alcock, C. et al. 2001, Nature, 414, 617

\bibitem[An et al.(2003)]{an02} An, J.H. et al. 2002, \apj, 549, 759

\bibitem[Ansari et al.(1999)]{ansari99} Ansari, R., et al. 1999, \aap, 344, L49

\bibitem[Aubourg et al.(1993)]{aubourg93} Aubourg, E. et al. 1993, 
Nature, 365, 623

\bibitem[Auri\`ere et al.(2001)]{auriere01} 
Auri\`ere, M., et al. 2001, \aap, 553, L137

\bibitem[Baillon et al.(1993)]{baillon93} Baillon, P., Bouquet, A.,
Giraud-H\'eraud, Y. \& Kaplan, J. 1993, \aap, 277, 1

\bibitem[Boden et al.(1998)]{BSV98}
Boden A. F., Shao M., \& Van Buren D. 1998, \apj, 502, 538

\bibitem[Bond(2004)]{bond04} Bond, I.A. et al. 2004, \apj, 606, L155

\bibitem[Calchi Novati(2003)]{calchi03} Calchi Novati, S. 2003, \aap, 405, 851

\bibitem[Castro et al.(2001)]{castro01} Castro, S., Pogge, R.W., Rich, R.M.,
DePoy, D.L., \& Gould, A. 2001, \apj, 548, L197

\bibitem[Cassan et al.(2004)]{cassan04} Cassan, A. et al. 2004, \aap, 419 L1

\bibitem[Crotts(1992)]{crotts92} Crotts, A.P.S. 1992, \apj, 399, L43

\bibitem[Dahn et al.(1995)]{dahn95} Dahn, C.C., Liebert, J.W., Harris, H., 
\& Guetter, H.C.\ 1995, 
p.\ 239, An ESO Workshop on: the Bottom of the Main Sequence and Beyond,
C.G.\ Tinney ed.\ (Heidelberg: Springer)

\bibitem[Drake et al.(2004)]{dck04}
Drake, A.J., Cook, K.H., \& Keller, S.C. 2004, \apj, 607, L29

\bibitem[de Jong et al.(2004)]{dejong04} de Jong et al. 2004, \aap, 417, 461

\bibitem[Joshi et al.(2005)]{joshi05} Joshi, Y.C., Pandey, A.K.,
Narasimha, D., \& Sagar, R. 2005, \aap, in press (astroph/0412550)

\bibitem[Fields et al.(2003)]{fields03} Fields, D.L. et al. 2003, 
\apj, 596, 1305

\bibitem[Gaudi et al.(2002)]{gaudi02} Gaudi, B.S. et al. 2002, \apj, 566, 463

\bibitem[Gould(1994)]{gould94}
Gould, A. 1994, \apjl, 421, L75

\bibitem[Gould(1995a)]{gould95a}
Gould, A. 1995a, \apjl, 441, L21

\bibitem[Gould(1995b)]{gould95b} Gould, A. 1995b, \apj, 441, 77

\bibitem[Gould(2000)]{gould00} Gould, A. 2000, \apj, 542, 785

\bibitem[Gould(2001)]{gould01} Gould, A. 2001, \pasp, 113, 903

\bibitem[Gould(2003)]{gould03} Gould, A., 2003, \apj, 583, 765

\bibitem[Gould(2004a)]{gould04a} Gould, A., 2004a, \apj, 606, 319

\bibitem[Gould(2004b)]{gould04b} Gould, A., 2004b, \apj, 607, 653

\bibitem[Gould et al.(2004)]{gba04}
Gould, A., Bennett, D.P., \& Alves, D.R., \apj, 614, 404

\bibitem[Gould et al.(1997)]{gbf97} Gould, A., Bahcall, J.N., \&
Flynn, C. 1997, \apj, 482, 913

\bibitem[Gould et al.(1998)]{gfb98} Gould, A., 
Flynn, C. Bahcall, \& J.N. 1998, \apj, 503, 798

\bibitem[Gould \& Loeb(1992)]{gould92} Gould, A. \& Loeb, A. 1992, 
\apj, 396, 104

\bibitem[Gould \& Salim(1999)]{gs99} Gould, A. \& Salim, S. 1999
\apj, 524, 794

\bibitem[Griest et al.(1991)]{griest91} Griest, K. et al. 1991, \apj, 372, L79

\bibitem[H{\o}g et al.(1995)]{HNP95}
H{\o}g, E., Novikov, I. D., \& Polanarev, A. G. 1995, \aap, 294, 287

\bibitem[Kiraga \& Paczy\'nski(1994)]{kiraga94} 
Kiraga, M. \& Paczy\'nski, B. 1994, \apj, 430, L101

\bibitem[Kubas et al.(2005)]{kubas05} Kubas, D. et al. 2005, \aap, in press
(astroph/0502018)

\bibitem[Mao \& Paczy\'nski(1991)]{mao91} Mao, S. \& Paczy\'nski, B. 1991, \apj, 
374, 37

\bibitem[Miyamoto \& Yoshii(1995)]{MY95}
Miyamoto,  M., \& Yoshii, Y. 1995, \aj, 110, 1427

\bibitem[Paczy\'nski(1986)]{pac86} Paczy\'nski, B. 1986, \apj, 304, 1

\bibitem[Paczy\'nski(1991)]{pac91} Paczy\'nski, B. 1991, \apj, 371, L63

\bibitem[Paczy\'nski(1998)]{Pa98}
Paczy\'nski, B. 1998, \apjl, 494, L23

\bibitem[Popowski(2005)]{popow05} Popowski, P. et al. 2005, \apj, submitted 
(astroph/0410319)

\bibitem[Refsdal(1966)]{Re66}
Refsdal, S. 1966, \mnras, 134, 315

\bibitem[Riffeser(2003)]{riffeser03} Riffeser, A., Fliri, J., Bender, R.,
Seitz, S., \& G\"ossl, C.A. 2003, \apj, 599, L17

\bibitem[Rhie et al.(2000)]{rhie00} Rhie, S.H. et al. 2000, \apj, 533, 378

\bibitem[Sumi et al.(2005)]{sumi05} Sumi, T. et al. 2005, \apj, submitted
(astroph/0502363)

\bibitem[Tisserand \& Milsztajn(2005)]{tisserand05} 
Tisserand, P. \& Milsztajn, A.
2005, Proceedings of the 5th Rencontres du Vietnam ``New Views on the
Universe'', in press, astroph/0501584

\bibitem[Udalski et al.(1993)]{udalski93} Udalski, A., Szymanski, M., 
Kaluzny, J., Kubiak, M., Krzeminski, W., Mateo, M., Preston, G.W., 
Paczy\'nski, B. 2003, Acta Astron., 43, 289 

\bibitem[Udalski(2003)]{udalski03} Udalski, A. 2003, Acta Astron., 53, 291 

\bibitem[Uglesich et al.(2004)]{uglesich04} Uglesich, R.R. et al. 
2004, \apj, 612, 877

\bibitem[Walker(1995)]{Wa95}
Walker, M. A. 1995, \apj, 453, 37


\end{thebibliography}
\end{document}